# Esaki diodes in van der Waals heterojunctions with broken-gap energy band alignment

*Rusen Yan[1,2*], Sara Fathipour[2], Yimo Han[4], Bo Song[1,2], Shudong Xiao[1], Mingda Li[1], Nan Ma[1],*

*Vladimir Protasenko[2], David A. Muller[4,5], Debdeep Jena[1,2,3], Huili Grace Xing[1,2,3*]*

1. School of Electrical and Computer Engineering, Cornell University, Ithaca, NY 14583

2. Department of Electrical Engineering, University of Notre Dame, Notre Dame, IN 46556

3. Department of Materials Science and Engineering, Cornell University, Ithaca, NY 14583

4. School of Applied and Engineering Physics, Cornell University, Ithaca, NY 14853

5. Kavli Institute at Cornell for Nanoscale Science, Ithaca, NY 14853

* ry253@cornell.edu; grace.xing@cornell.edu

**Van der Waals (vdW) heterojunctions composed of 2-dimensional (2D) layered materials are emerging as a solid-state materials family that exhibit novel physics phenomena that can power a range of electronic and photonic applications[1,2]. Here, we present the first demonstration of an important building block in vdW solids: room temperature Esaki tunnel diodes. The Esaki diodes were realized in vdW heterostructures made of black phosphorus (BP) and tin diselenide ($SnSe_2$), two layered semiconductors that possess a broken-gap energy band offset. The presence of a thin insulating barrier between BP and $SnSe_2$ enabled the observation of a prominent negative differential resistance (NDR) region in the forward-bias current-voltage characteristics, with a peak to valley ratio of 1.8 at 300 K and 2.8 at 80 K. A weak temperature dependence of the NDR indicates electron tunneling being the dominant transport mechanism, and a theoretical model shows excellent agreement with the experimental results. Furthermore, the broken-gap band alignment is confirmed by the junction photoresponse and the phosphorus double planes in a single layer of BP are resolved in transmission electron microscopy (TEM) for the first time. Our results**

**represent a significant advance in the fundamental understanding of vdW heterojunctions, and broaden the potential applications of 2D layered materials.**

Esaki's discovery of NDR in heavily doped semiconducting germanium *p-n* junctions in 1958 was the first experimental evidence of quantum mechanical tunneling transport of electrons in all-condensed-matter systems[3,4]. This discovery motivated Giaever's tunneling experiments that proved the existence of the superconductive energy gap predicted by the then-newly formulated Bardeen-Cooper Schrieffer (BCS) theory of superconductivity[5]. After these initial breakthroughs, tunneling in various classes of crystalline matter has been observed, and forms the basis for several practical applications. For example, Josephson junctions exploit tunneling in superconductors for exquisitely sensitive magnetic flux detectors in superconducting quantum interference devices (SQUIDs)[6], and are now being investigated as the building blocks of quantum computers[7]. Electron tunneling forms the basis for low-resistance ohmic contacts to heavily doped semiconductors for energy-efficient transistors, as low-loss cascade elements in multi-junction solar cells, and for coherent emission of long-wavelength photons in quantum-cascade lasers[8,9]. In addition to such practical applications, the extreme sensitivity of tunneling currents to various electronic, vibrational, and photonic excitations of solids makes tunneling spectroscopy one of the most sensitive probes for such phenomena[10].

Recently, interband tunneling in semiconductors has been proposed as the enabler for a new class of semiconductor transistors called tunnel field-effect transistors (TFETs) that promise very low-power operation. The heart of such devices is an Esaki tunnel diode, with preferably a near broken-gap band alignment at the source-channel heterojunction[11,12]. As these heterojunction TFETs are scaled down to the nanometer regime, the increase in bandgap barrier due to quantum confinement may significantly prohibit the desired tunneling currents, because tunneling current decreases exponentially with the barrier height. Layered semiconductors with a sizable bandgap and a wide range of band alignments can potentially avoid such degradation, and have been proposed as ideally suited for such applications[13,14]. This class of devices distinguishes themselves from the graphene-based SymFET by offering a desired low off-current[15,16]. Compared to traditional 3D heterojunctions, such structures are expected to form high-quality heterointerfaces due to the absence of dangling bonds[1–3,20]. The weak vdW bonding in principle does not suffer from lattice mismatch requirements and makes strain-free integration possible. Among the previous reports on vdW solids, heterojunctions of type-I (straddling) and type-II (staggered) band alignments have been demonstrated[17–19]. Very recently, Roy et al. reported an Esaki NDR at low temperatures in an as-stacked $MoS_2/WSe_2$ heterojunction which is believed to possess a type-II band alignment, employing dual gates.[20] However, the NDR observed by them did not persist to room temperature. In this work, two layered materials – BP and $SnSe_2$ - are successfully integrated for the first time, enabling the conclusive achievement of Esaki-diode behavior in 2D crystal semiconductors at room temperature. We use the Esaki diodes to experimentally prove that the heterojunction possesses a type-III (broken) band alignment. The BP/SnSe$_2$ heterojunction is one between *mixed-valence* materials, because black phosphorus is elemental and SnSe$_2$ is a compound semiconductor. Furthermore, the two constituents have different *crystal structures*. Given all these differences, it is rather remarkable that a robust NDR can be observed at room temperature in this heterostructure.

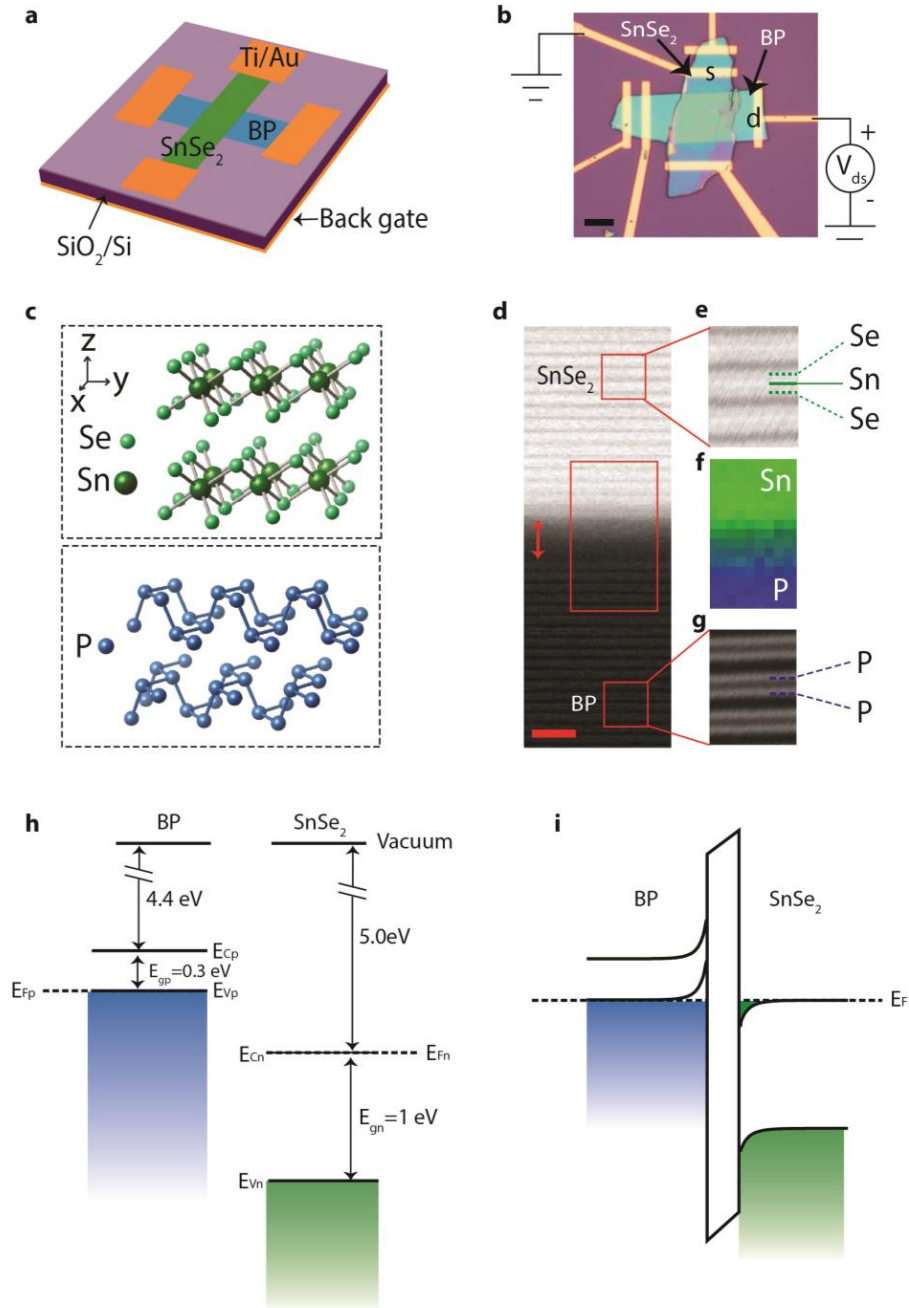

*Figure 1. Structure and energy band alignment of the BP/SnSe$_2$ vdW heterojunction.* ***a*** *and **b**,*

*Schematic illustration and optical image of fabricated devices on SiO$_2$/Si substrates. The scale bar is 5 um. A voltage V$_{ds}$ is applied on the p-type BP with the n-type SnSe$_2$ grounded. **c**, Crystal structures of BP and SnSe$_2$. BP is composed of two planes of phosphorus atoms arranged in puckered layers and SnSe$_2$ has a CdI$_2$ type structure with each plane of Sn atoms sandwiched between two planes of Se atoms. Layers in BP or SnSe$_2$ are bonded together by vdW force, which allows exfoliation of these crystals into thin flakes. **d, e,** and **g**, Cross-sectional STEM images. **d** shows the presence of a thin barrier (~ 1.6 nm) at the interface. The scale bar is 2 nm. **e** shows the tri-atomic-plane layer structure of SnSe$_2$. **g** shows the bi-atomic-plane layer structure of BP. The measured interlayer distance of SnSe$_2$ is 0.65 nm and that of BP is 0.55 nm. **f**. EELS map: Sn in green and P in blue. **h**, Energy band profiles of BP and SnSe$_2$ prior to contacting each other[21,22]. BP has a low work function compared to that of SnSe$_2$. Due to un-intentional doping, exfoliated BP and SnSe$_2$ flakes onto SiO$_2$/Si substrates are p-type[23] and heavily n-type doped, respectively (also see supplementary Fig. S1). **i,** Band alignment at equilibrium. Accumulations of holes in BP and electrons in SnSe$_2$ result from electron transfer from BP to SnSe$_2$ owing to a lower work function of BP than that of SnSe$_2$. The heavily degenerately doped p-BP and n-SnSe$_2$ along with the thin insulating barrier in between form the heart of the Esaki diode.*

The Esaki diode device structure is schematically shown in Fig. 1a. The devices are fabricated using a dry transfer process with flake thicknesses of ~50-100 nm for both BP and SnSe$_2$[24]. BP is the p-type semiconductor, and SnSe$_2$ the n-type semiconductor of the vdW Esaki diode. A detailed device fabrication and characterization procedure is provided in the Method section. Figure 1b shows an optical image of a representative device. The thicknesses of BP and SnSe$_2$ in this device are 79 and 95 nm, determined by atomic force microscopy (AFM). Figure 1c shows the crystal structures of BP and SnSe$_2$. BP, the most stable allotrope of phosphorus obtained under high pressure, possesses an orthorhombic crystal structure. Each phosphorus atom is covalently bonded to three adjacent atoms in a puckered layer, thus forming two atomic planes in each layer of BP[25]. In comparison, SnSe$_2$ consists of planes of Sn sandwiched between two planes of Se atoms, forming a hexagonal CdI$_2$-type structure[26]. Similar to other 2D crystals, layers are held together by vdW forces in both BP and SnSe$_2$. Figure 2d shows the high-angle annular dark-field scanning transmission electron microscopy (HAADF-STEM) images of the SnSe$_2$-BP interface. Figure 2e and 2g are the zoomed-in HAADF-STEM images of SnSe$_2$ and BP, showing the tri-atomic-plane SnSe$_2$ layers with an interlayer distance of 0.65 nm and the bi-atomic-plane BP layers with an interlayer separation of 0.55 nm. The measured layer thicknesses for both materials are very close to the values in the literature. At the interface, an amorphous layer with a thickness ranging from 1.2 nm to 2 nm is observed, which is thicker than a typical vdW gap of 0.4 – 0.6 nm[20,27]. Electron energy loss spectroscopy (EELS) composition analysis of this layer is shown in Fig. 2f. It reveals carbon and traces of phosphorus and Sn (also see supplementary Fig. S3). We speculate that the presence of carbon most likely stems from the flake transfer process, while the presence of P and Sn likely results from the finite interdiffusion, assisted by the degradation of the BP and SnSe$_2$ flake surfaces prior to being stacked together due to their widely known instability in air[28,29]. We can not completely rule out that BP and SnSe$_2$ near the interface suffered from degradation during the TEM sample preparation and imaging process, as observed by other groups[27]. However, it is worth highlighting that the double atomic planes in a BP layer is resolved by TEM for the first time in this work, to the best of our knowledge. This can be attributed to the aberration correction up to the 5$^{th}$ order and the low electron beam energy used: 100 keV. More details on STEM can be found in the Method section.

High doping densities boost the NDR characteristics in traditional semiconductor Esaki diodes, which applies in these vdW Esaki diodes as well. The BP flake is unintentionally p-type doped, and the SnSe$_2$ flake is unintentionally doped n-type. This is consistent with other groups' observations[23,30]. We confirm the individual doping types unambiguously by the opposite directions of field-effect conductivity modulation of each layer (supplementary Fig. S1a). That the effective doping densities are high manifests in the relatively weak current modulation in the individual flakes, which is also attributed to the large flake thicknesses. Figure 1h shows the energy level alignments of conduction and valence band edges of the *p*-type BP and *n*-type SnSe$_2$ based on the reported electron affinity values in the literature[21,26]. Based on these alignments, they are expected to form a type-III, broken-gap heterojunction similar to InAs/GaSb tetrahedral 3D semiconductors. The large work function difference leads to an accumulation of holes in BP and electrons in SnSe$_2$ near the junction, and an effective *p-i-n* junction is formed in our devices owing to the presence of the aforementioned interfacial layer (Fig. 1i). Away from the junction, it is assumed that the unintentional *p*-type doping is such that it puts the Fermi level of p-BP ($E_{fp}$) near its valence band edge ($E_{vp}$), and the doping in *n*-SnSe$_2$ puts its $E_{fn}$ near its conduction band edge ($E_{cn}$). Both layers are thus effectively 'degenerately' doped. This junction, made of a heavily doped *p* region and a heavily doped *n*- region separated by a thin tunneling barrier forms the heart of the Esaki diode, enabling the observation of NDR, as will be discussed in the following text. Though the tunnel barrier in this work is formed unintentionally, thin BN layers or other suitable barrier materials with improved quality[27] can be used: which is the focus of our future work.

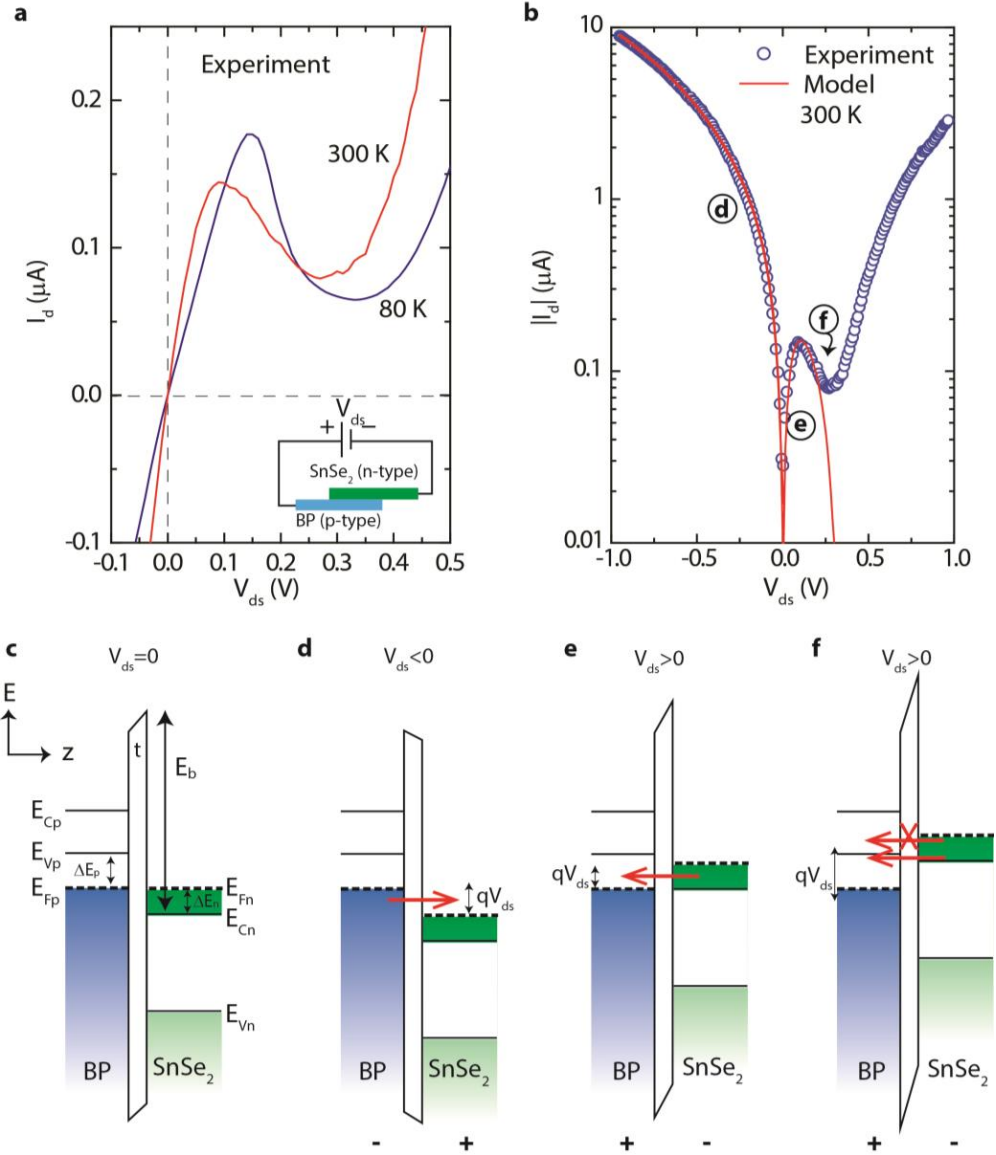

*Figure 2. Current-voltage characteristics of the BP/SnSe₂ vdW Esaki tunnel diode.* ***a***, $I_d$-$V_{ds}$ curves at 80 and 300 K in a linear scale. At lower temperatures, the peak current increases and the valley current decreases thus an increase in the peak-to-valley ratio, as expected from the theory and consistent with the prior reported Esaki diodes.[4] ***b***, Comparison of experimental and theoretical $I_d$-$V_{ds}$ curves at 300 K in a log scale. Excellent agreement is achieved. ***c, d, e, f,*** Band alignment at $V_{ds}$=0, $V_{ds}$<0, and $V_{ds}$>0. When the conduction band electrons in SnSe₂ have a maximal overlap with the valence band holes in BP ($V_{ds}$>0), the Esaki tunnel diode reaches its peak current. With further increasing $V_{ds}$ (***f***), part of the conduction band electrons in SnSe₂ see the forbidden band gap of BP, thus leading to a reduction in current.

Figure 2a presents the measured $I_d$-$V_{ds}$ curves at 80 and 300 K in a linear scale, where the most notable feature is the appearance of NDR while the junction is positively biased ($V_{ds}$>0). The same data at RT are plotted in a logarithmic scale in Fig. 2b (blue circles) to showcase the backward diode behavior: current under reverse bias is higher than that at forward bias, opposite to a typical p-n diode. The simultaneous appearance of NDR and backward-diode behavior is a conclusive fingerprint of interband tunneling transport of electrons. All results presented in this work are obtained by applying a drain voltage $V_{ds}$ on BP (p-type, terminal "*d*") with the SnSe$_2$ (n-type, terminal "*s*") being grounded. For example, for $V_{ds}$>0 electrons drift through the conduction band of *n*-type SnSe$_2$, tunnel through the barrier into the empty valence band states in *p*-type BP, and drift through this valence band to emerge at the BP ohmic contact (see Figs 1b and 2e,f). The band alignments of the vdW Esaki diode close to the BP/SnSe$_2$ interface under various bias conditions are sketched in Fig. 2c-f. At equilibrium, the Fermi level $E_f$ lies between $E_{vp}$ and $E_{cn}$ (also see Fig. 1i). When a reverse bias voltage $V$ is applied (Fig. 2d), a finite energy window is created for electrons to tunnel from the filled valence band states of BP into the empty conduction band states in SnSe$_2$: the transport is driven by the difference between the hole quasi Fermi level $E_{fp}$ and electron quasi Fermi level $E_{fn}$ that satisfy: $E_{fp} - E_{fn} = qV$, where $q$ is the elementary charge. Increasing the reverse bias simultaneously widens the energy window for tunneling, and increases the tunneling probability because of a stronger field, leading to a sharp increase in the tunneling current. This is clearly observed in our devices, the region "d" in Fig 2b, and captured accurately with a tunneling transport model as described later. Under a small forward bias (0<$qV_{ds}$<$\Delta E_n$, where $\Delta E_n = E_{fn} - E_{cn}$), electrons residing in the conduction band of *n*-SnSe$_2$ are able to tunnel to the empty states in the valence band of *p*-BP through the energy window of $E_{fn} - E_{fp}$ (Fig. 2e). This tunneling current reaches its peak when the occupied conduction band states in SnSe$_2$ have a maximal overlap with the unoccupied valence band states in BP. Further increase of the voltage aligns the occupied conduction band states with the forbidden bandgap of BP. Then the tunneling current decreases in spite of a slight increase in tunneling probability induced by a stronger electric field across the barrier. The diode current reaches its minimum (valley) value, which is typically dominated by a combination of defect and phonon assisted tunneling and thermionic currents[31], followed by an upturn due to the overriding thermionic emission current.

A simple model of interband tunneling using the Wentzel-Kramers-Brillouin (WKB) approximation is able to capture all the essential experimental tunneling features quantitatively in these vdW heterojunction Esaki diodes. While the details of the model appear in the Supplementary Information (SI), its essence is described briefly here. The device tunneling current is calculated by summing the individual single-particle contribution from all electron states in the *k*-space[15] that are allowed to tunnel:

$$I_t = q \frac{g_s g_v}{L_z} \sum_{k_n} \sum_{k_p} v_z (F_n - F_p) T_{WKB} \delta(E_p - E_n) \delta(k_{xp} - k_{xn}) \delta(k_{yp} - k_{yn}) \quad (1)$$

where $g_s$ and $g_v$ are the spin and valley degeneracy in the bandstructure, $L_z$ is the macroscopic device length along the electric field direction $z$, $v_z = \hbar k_{zn}/m_{zn}$ is the band group velocity in the source side, $F_n$ and $F_p$ are the Fermi-Dirac distribution functions, respectively, in $n$-SnSe$_2$ and $p$-BP, and $T_{WKB}$ is the WKB tunneling probability. The three Dirac-$\delta$ functions in equation (1) simultaneously ensure energy ($E$) conservation and in-plane momentum ($k_x$ and $k_y$) conservation. The numerically evaluated tunneling current is shown in Fig. 2b as a red solid line, and is seen to explain the entire tunneling portion of the current: both in reverse and forward bias, before the thermionic/diffusive-part takes over at large forward bias. The model incorporates an effective barrier lowering of $\Delta\phi_b = \sqrt{q\Xi/4\pi\epsilon_r\epsilon_0}$ where a relative permittivity $\epsilon = 6$ is used and $\Xi$ is the electric field across the tunnel barrier. At 300 K, the best fit of the model to the experimental tunneling current is obtained by setting $\Delta E_p = E_{vp} - E_{fp} = 0.27$ eV, $\Delta E_n = E_{fn} - E_{cn} = 0.07$ eV, a barrier height $E_b = 4.7$ eV, and a tunneling barrier width $t=1.3$ nm. The effective doping induced degeneracies and energy scales are reasonable, and the tunnel barrier thickness is in agreement with the physical barrier thickness measured by STEM for the vdW $p+/i/n+$ heterointerface. Recognizing that the tunneling probability is exponentially dependent on the barrier height and thickness, a peak tunnel current map is generated as a function of these two parameters in Fig. 3. For example, the experimentally observed peak current at RT can be reproduced using a barrier thickness of 2 nm with a barrier height of 2 eV as well as a barrier thickness of 1.3 nm with a barrier height of 4.7 eV. Considering the effective tunneling area might be smaller than the physical overlap region area (~10x10 μm$^2$) due to the potential surface contamination during dry transfer, we believe the property of the tunnel barrier between BP and SnSe$_2$ is most likely spatially distributed and largely falls in the parameter window explored in Fig.3. A number of BP/SnSe$_2$ vdW diodes have been fabricated, among which several showed stable NDR behavior under forward bias, while some showed backward diode behavior without NDR. The NDR in those samples is washed out by overwhelmingly high valley currents under forward bias. More information is provided in supplementary Fig. S4 and S5.

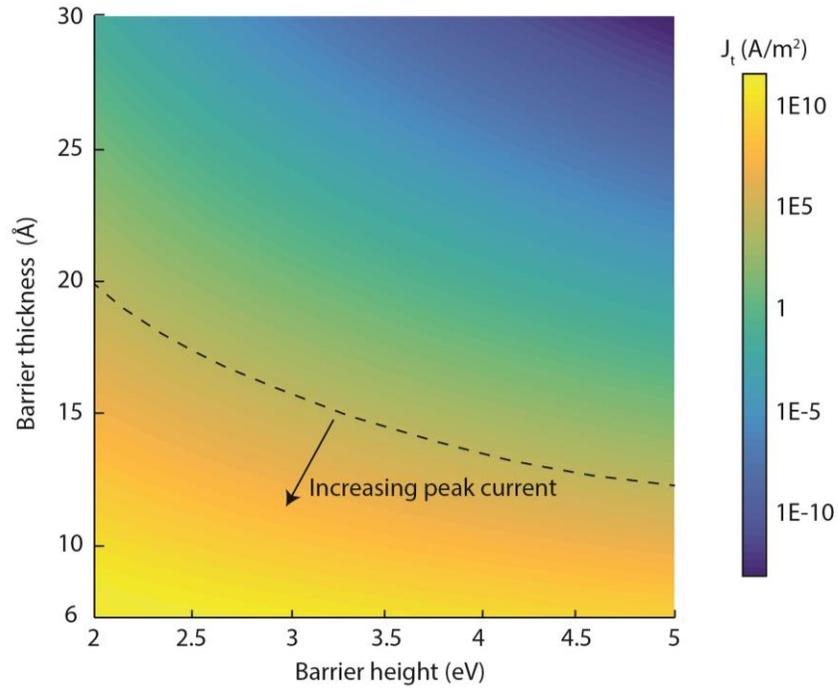

*Figure 3. Calculated peak current density in the BP/SnSe$_2$ Esaki diode mapped over a range of tunnel barrier heights and thicknesses.* The dash line marks the measured peak current density shown in Fig.2, which is around 1.6x10$^3$ A/m$^2$. For the modeled $I_d$-$V_{ds}$ curve shown in Fig. 2, the following parameters: $E_b$=4.7 eV and t=13 Å, render an excellent match with the measured $I_d$-$V_{ds}$.

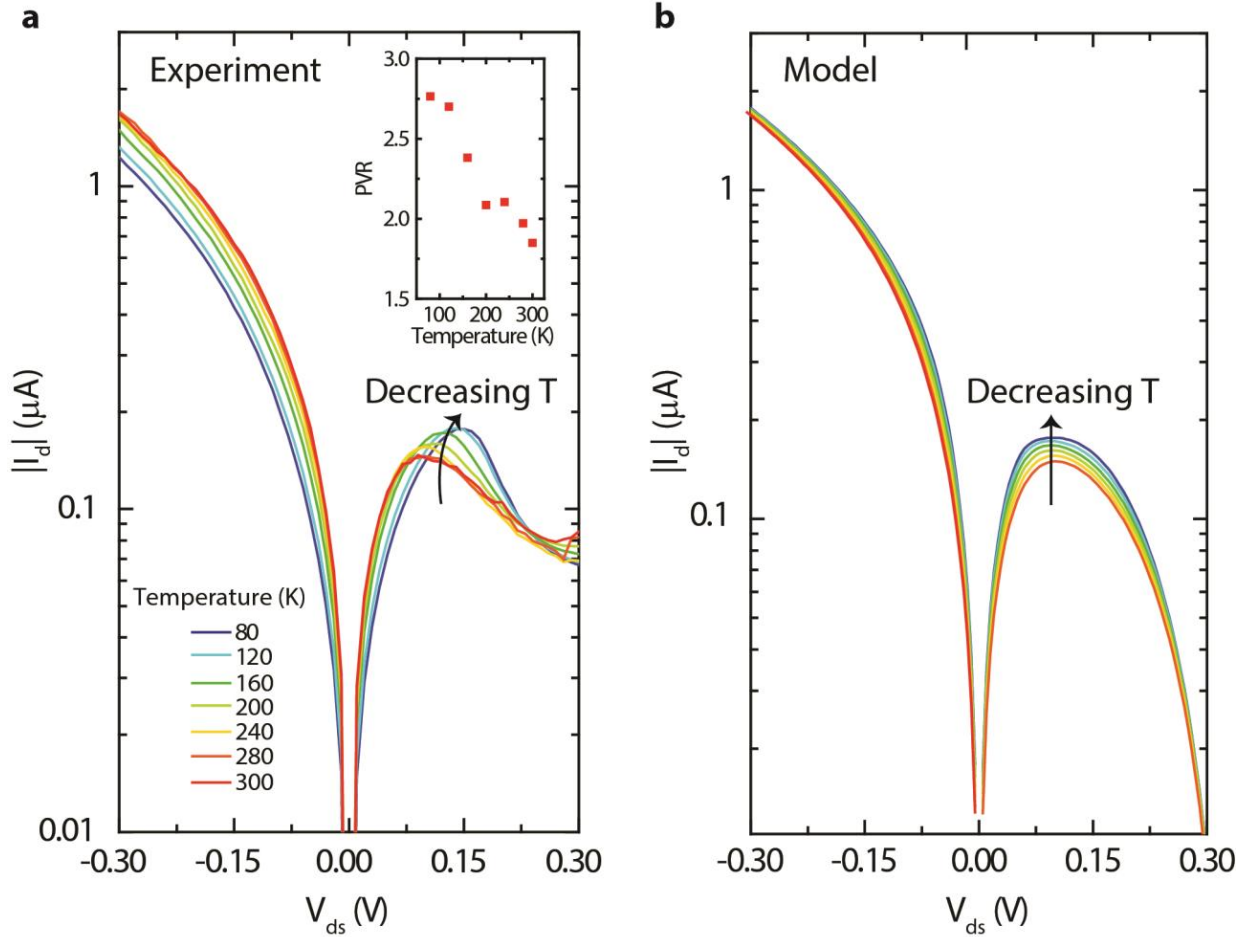

*Figure 4. Temperature dependent current-voltage characteristics of the BP/SnSe₂ Esaki diode. a, b, Experimental and theoretical $I_d$-$V_{ds}$ curves in a temperature range of 80 to 300 K. As temperature decreases, the peak current increases as a result of a tighter Fermi-Dirac distribution of the carriers. The devices suffered from poor ohmic contacts at low temperatures, which shifts the peak voltage to a higher value under the forward bias and lowers the apparent tunneling current under the reverse bias. Inset of a shows the measured peak to valley ratio as a function of temperature.*

The current contribution from tunneling mechanism is expected to have a weak dependence on temperature to the extent of thermal smearing of the Fermi occupation functions. To verify this, we performed temperature dependence $I_d$-$V_{ds}$ measurements. Figure 3 shows the measured and modeled $I_d$-$V_{ds}$ curves for temperatures ranging from 80 to 300 K. The measured temperature dependence is in excellent agreement with the model, highlighting its predictive capability in spite of the simplicity. A critical feature in the characteristics is the evolution of the peak current $I_p$ in the NDR region as a function of temperature: $I_p$ increases slightly with decreasing *T*. The Fermi-Dirac tail shortens with decreasing temperature, making the electron occupation probability difference in the *n*- and *p*- sides $F_n - F_p$ larger, increasing $I_p$. This temperature dependence is in stark contrast to thermionic emission, or phonon/trap-assisted tunneling, which typically shows an opposite trend of decreasing current with decreasing temperature. The peak voltage is observed to shift to a higher value at lower temperatures because the ohmic contacts confirmed at RT become Schottky-like at lower temperatures (see supplementary Fig. S2a/b). For the same reason, the reverse bias currents are seen to be slightly lower at low T than that at RT. The inset of Fig. 3a shows that the extracted peak-to-valley ratio (PVR) of NDR varies from 2.8 at 80 K to 1.8 at 300K. Similar trends are also observed in Esaki diodes made of 3D semiconductors, such as Ge, Si[3,4].

The broken-gap energy band alignment of the BP/SnSe$_2$ vdW interface is unusual, and of high interest for future applications. The tunneling current model provides the first sign of this band alignment. To conclusively verify the broken-gap energy alignment, we explore the photo response of the heterojunction under illumination from a 488 nm laser, as shown in Fig 5. Since the Esaki diode forms a perfect "ohmic" contact near zero voltage, the sign of the photocurrent and photovoltage uniquely correlates with the energy band bending in the heavily doped $p+$ and $n+$ regions near the junction. If the band bending is such that carriers of opposite sign *accumulate* near the junction as shown in the inset of Fig. 5a, the resulting photocurrent will move the Esaki diode I-V curve up into the 2$^{nd}$ quadrant. If the band bending is such that carriers of opposite sign *deplete* near the junction as a typical p-n diode, the photoresponse I-V will move into the 4$^{th}$ quadrant, like a solar cell. Provided that the Fermi level in *p*-BP (*n*-SnSe$_2$) is very close to its valence (conduction) band edge as argued above, the movement of the I-V curve into the 2$^{nd}$ quadrant in Fig 5a, and the resulting negative open-circuit voltage is a signature of a broken-gap alignment. Though it is impossible to rule out the formation of unintentional charged states at the vdW heterointerface in this study, the energy band alignment in BP/SnSe$_2$ diode is most likely of type-III. The observed Esaki diode behavior and its peculiar photo response, combined with the reported properties of the layered materials point strongly towards this conclusion. The detailed discussion on various band alignment possibilities can be found in the SI. With an absorption coefficient of ~5×10$^5$ cm$^{-1}$ of SnSe$_2$ bulk crystals at 488 nm[32], we estimate the laser power reaching the vdW heterojunction, and the device responsivity at zero bias to be ~0.24 mA/W. This value is comparable to those reported in other layered materials[21].

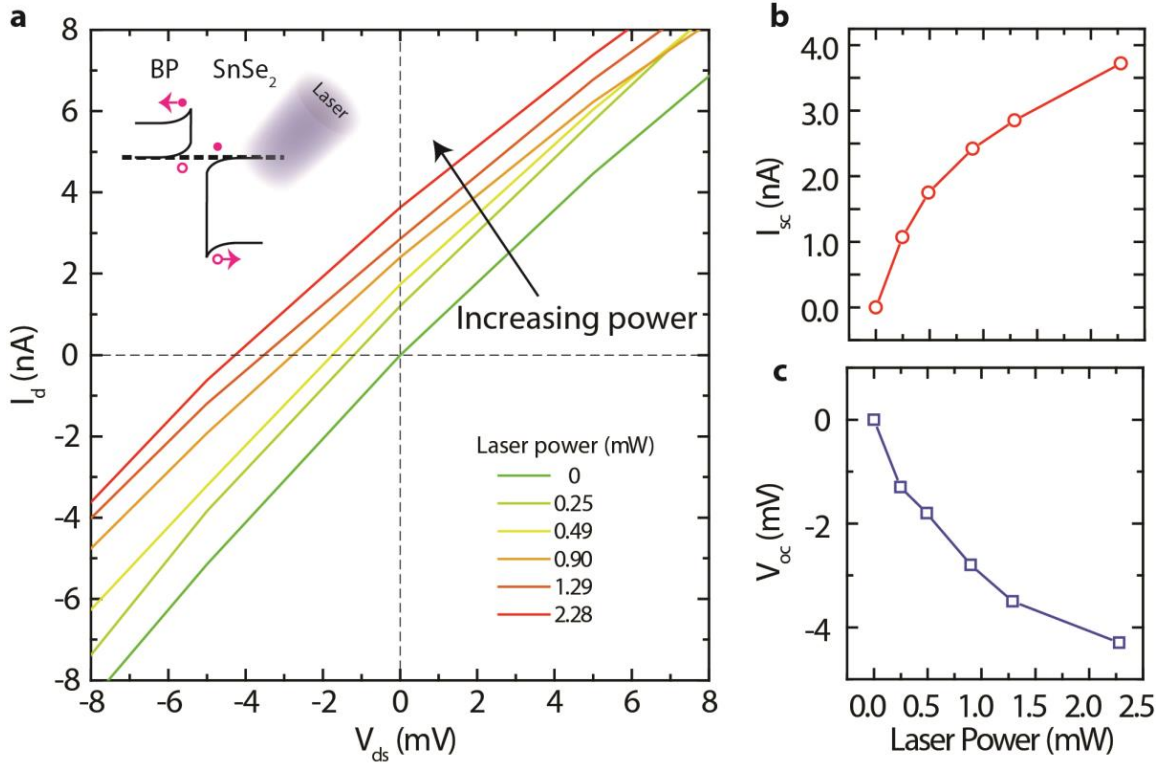

*Figure 5. Current-voltage behavior of the BP/SnSe$_2$ Esaki diode under illumination of a 488 nm laser. a, $I_d$-$V_{ds}$ curves near zero bias. Laser power is varied from 0 to 2.28 mW and the laser beam diameter is about 3 um illuminating the vdW junction only, excluding the metal/2D regions. The sign of the short-circuit current $I_{sc}$ and the open-circuit voltage $V_{oc}$ is opposite to that of a common p-n junction or solar cell, confirming the band bending in BP (accumulation of holes) and SnSe$_2$ (accumulation of electrons) thus the type-III or broken band alignment. The inset shows this confirmed band bending. b, c, $I_{sc}$ and $V_{oc}$ as a function of the laser power.*

In conclusion, we have demonstrated an Esaki tunnel diode using vdW heterojunctions with a type-III energy band alignment composed of BP and SnSe$_2$ for the first time. Our study suggests 2D crystals as promising candidates for future tunneling based devices. Finally, we point out that Esaki et al. observed NDR in metal-insulator-SnTe junctions, where they realized that the degenerate p-type doping of SnTe coupled with the modulation of the tunneling probability with voltage was primarily responsible for the observed NDR[4]. In that study, they realize the crucial role played by the fact that the band alignment of the metal and the semiconductor are not "locked" as in epitaxial III-V heterostructures, but can slide with respect to each other because of the presence of the insulating barrier. They call this a new type of tunneling, which indeed is identical to what we see in the BP/SnSe$_2$ layered semiconductor heterojunction. However, this story has an even older precedent – in Holm and Meissner's experiments in 1930's, where they attributed the resistance measured between two metals to tunneling of electrons through a 'void' between them[33]. In the layered heterojunction materials, the lack of strong interlayer covalent bonds presents exactly such an electronic void, which helps decouple the band alignment and allows the robust NDR observed here. On a more general note, we mention that the robust Esaki-type NDR reported in this work is not entirely new – and has been seen over the past half century in various group-IV and III-V semiconductor heterostructures. However, its observation in layered semiconductors that lack out of plane chemical bonds is potentially a starting point for several investigations in the future. Unlike III-V and group-IV semiconductors and their heterostructures, layered semiconductors can be semiconducting, insulating (BN), but also metallic and/or or superconducting (NbSe$_2$). This means one can create seamless heterostructures of semiconductors with superconductors and insulators without interface traps, and investigate the tunneling of particles - be it of single-particle type (in semiconductors), or correlated (in superconductors) between various material families. The fine structures in such tunneling experiments can also serve as the most sensitive tools for the measurement of band alignments, phonon modes, elastic vs inelastic tunneling physics, and a host of potentially unforeseen physics.

**Method**

**Device Fabrication.** The device fabrication starts with the mechanical exfoliation of BP flakes (Smart-element GmbH) onto a 285 nm thick SiO$_2$ covered silicon substrate using scotch tape. The 2nd layered material SnSe$_2$ (2D Semiconductor Inc.) is first exfoliated onto commercially available elastomeric films as stamps. The films (Gel-Fim® WF x4 6.0mil) were supplied by Gel-Pak. After locating the appropriate thin flakes on the stamp film, the film is turned upside down and aligned with the target BP flake using micro-manipulators under an optical microscope. Once the desired alignment is achieved, the stamp is pressed against the substrate and then removed slowly to ensure that the SnSe$_2$ flake remains adhered to the BP flake. This all-dry transfer process avoids liquid contamination and can be completed in a reasonably short amount of time, both of which are critical for fabricating high quality devices.[24] 10/100 nm thick Ti/Au metal contacts are defined by electron beam lithography and a lift-off process. A Tl/Au stack was deposited on the backside of the Si substrate to serve as the back-gate.

**Photocurrent Measurement.** Photocurrents were measured with a Keithley 2636A source meter under optical excitation from a laser of wavelength 488 nm using an 8x objective lens with NA of 0.2. The laser spot size is around 1.22$\lambda$/NA~3 $\mu$m, which is much smaller than the size of the heterostructure overlapping area between BP and SnSe$_2$ (~10×10 $\mu m^2$).

**STEM specimen preparation and imaging.** A cross-sectional specimen of the device was prepared by using a standard lift-out procedure in a dual-beam FEI Strata 400 focus ion beam (FIB) system, with a final milling at 2 keV. After removal from the FIB system, the specimen was directly transferred into an ultrahigh vacuum bake-out system. The specimen was baked in the ultrahigh vacuum chamber for 8 hours at 130°C to clean the specimen before loading into the high vacuum of the electron microscope. During this procedure, the specimen experienced an approximately 1 minute long exposure to air. HAADF-STEM images were taken with a Nion Ultra-STEM 100 operated at 100 kV. The imaging condition was similar to that in Ref.34. The beam convergence angle was 35 mrad, with a probe current of ~70 pA. The acquisition time was 32 μs per pixel.

The EELS spectrum and images were acquired with an energy dispersion of 0.5 eV/channel using a Gatan Quefina dual-EELS Spectrometer. The beam convergence angle was 35 mrad, with a probe current of ~130 pA. A linear combination of power laws (LCPL) was used to fit and subtract the background. The EELS false-color composition map was created by integrating the Sn-$M_{4,5}$ edge and P-$L_{2,3}$ edge. Here we use the Sn-$M_{4,5}$ edge since the Se-$M_{4,5}$ edge was overwhelmed by a plasma peak on the BP side. All the EELS analysis was done with open-source Cornell Spectrum Imager software.[35]

**Author Contributions**

R. Y., D. J. and H. G. X. conceived experiments and conducted device modeling. R. Y. performed device fabrication, electrical and optoelectronic measurements, and data analysis. R. Y., B. S. and S. F. conducted low temperature measurements. V. P. helped on optical measurements. Y. H. took the HAADF-STEM images and conducted EELS spectrum analysis under supervision of D. A. M.. S. X., M. L. and N. M. helped with data analysis. R. Y., D. J. and H. G. X. wrote the manuscript. All authors have given approval to the final version of the manuscript.

**Acknowledgement**

This work was supported in part by NSF CARRER, NSF, AFOSR, and the Center for Low Energy Systems Technology (LEAST), one of the six SRC STARnet Centers, sponsored by MARCO and DARPA.. The device fabrication is performed at Cornell Nano-Scale Science and Technology Facility, funded by National Science Foundation. This work made use of the electron microscopy facility of the Cornell Center for Materials Research (CCMR) with support from the National Science Foundation Materials Research Science and Engineering Centers (MRSEC) program (DMR 1120296). We thank Prof. Alan Seabaugh, Dr. Pei Zhao, Malcolm Thomas and Megan Holtz for useful discussions.

*Supplementary information for*

# Esaki diodes in van der Waals heterojunctions with broken-gap energy band alignment

*Rusen Yan[1,2*], Sara Fathipour[2], Yimo Han[4], Bo Song[1,2], Shudong Xiao[1], Mingda Li[1], Nan Ma[1], Vladimir Protasenk[2], David Muller[4,5], Debdeep Jena[1,2,3], Huili Grace Xing[1,2,3*]*

1. School of Electrical and Computer Engineering, Cornell University, Ithaca, NY 14583

2. Department of Electrical Engineering, University of Notre Dame, Notre Dame, IN 46556

3. Department of Materials Science and Engineering, Cornell University, Ithaca, NY 14583

4. School of Applied and Engineering Physics, Cornell University, Ithaca, NY 14853

5. Kavli Institute at Cornell for Nanoscale Science, Ithaca, NY 14853

* ry253@cornell.edu; grace.xing@cornell.edu

**Contents:**

1) **Derivation of tunnel current in BP/SnSe$_2$ heterostructures**

2) **Additional electrical and TEM characterizations of the reported device**

3) **Reproducibility of vdW Esaki diodes with NDR**

4) **Observation of backward diode behavior without NDR**

5) **Deduction of band alignment from photocurrent measurements**

1) **Derivation of tunnel current in BP/SnSe₂ heterostructures**

The tunneling current is obtained by summing the individual contributions from each electron state in the *k*-space. The allowed transitions are governed by energy conservation and and lateral momentum conservation. The tunneling current is[1],

$$I_t = q \frac{g_s g_v}{L_z} \sum_{k_n} \sum_{k_p} v_z (F_p - F_n) T_{WKB} \delta(E_p - E_n) \delta(k_{xp} - k_{xn}) \delta(k_{yp} - k_{yn}) \quad (S1)$$

where $g_s=2$ is the spin degeneracy and $g_v$ the valley degeneracy. $L_z$ is the macroscopic length along the electric field direction *z*, $v_z$ is the electron group velocity in the tunneling direction. $F_n$ and $F_p$ are the Fermi-Dirac distribution functions of carriers in the *n* and *p* type materials respectively. The presence of the $\delta$ functions ensures energy and lateral momentum conservation, selecting only those states that are allowed to tunnel. $T_{WKB}$ is the WKB tunneling probability term. We convert the sum over crystal momentum $\sum_k (...)$ into the integral $L_x L_y L_z / (2\pi)^3 \times \int dk_x dk_y dk_z (...)$, and obtain the current density

$$J_t = \frac{I_t}{L_x L_y} = q g_s g_v \frac{V_p}{(2\pi)^6} \iiint dk_{nx} dk_{ny} dk_{nz} v_{nz} \iiint dk_{px} dk_{py} dk_{pz} v_{nz} (F_n - F_p) T_{WKB}$$

$$\delta(E_p - E_n) \delta(k_{xp} - k_{xn}) \delta(k_{yp} - k_{yn}) \quad (S2)$$

where $V_p$ is the volume of *p* type material. The total energy of electrons in *n* and *p* side are

$$E_n = E_{cn} + E_{kn} = E_{cn} + \frac{\hbar^2 k_{xn}^2}{2m_{xn}} + \frac{\hbar^2 k_{yn}^2}{2m_{yn}} + \frac{\hbar^2 k_{zn}^2}{2m_{zn}} \quad (S3),$$

$$E_p = E_{vp} - E_{kp} = E_{vp} - \frac{\hbar^2 k_{xp}^2}{2m_{xp}} - \frac{\hbar^2 k_{yp}^2}{2m_{yp}} - \frac{\hbar^2 k_{zp}^2}{2m_{zp}} \quad (S4),$$

where the conduction band edge in *n* is $E_{cn}$, the valence band edge in *p* is $E_{vp}$, and $E_k$ is the carrier kinetic energy. Considering the high effective masss anisotropy of BP and SnSe₂, we separate the carrier effective mass in the three directions. The tunneling probability is estimated using $T_{WKB} = \exp(-2\int_0^t |\kappa_z| dz)$. The imaginary wave vector $\kappa_z$ inside the barrier is given by

$$|\kappa_{bz}| = \sqrt{\frac{2m_b^* [E_b - \frac{q(V+V_i)}{t} z - (E_{cn} + E_{kn})]}{\hbar^2}} = \sqrt{\frac{2m_b^* (E_b - E_{cn} - E_{kn} - \frac{q(V+V_i)}{t} z)}{\hbar^2}} \quad (S5)$$

where *t* is the tunneling barrier width, $m_b^*$ is the effective mass of electrons inside the barrier, and $E_b$ is the energy difference from the conduction band edge of n-type material to the top of the barrier. It is difficult for us to determine these numbers uniquely based on prior reports in the literature. So we have set them as fitting parameters, which would be obtained through the best fit of the model to the experimental results. In the above expression, $V_i$ is the voltage drop over the barrier at zero bias induced by charge accumulation adjacent to the barrier.

In the absence of phonon scattering, energy conservation rule and $k_{xn} = k_{xp}$, $k_{yn} = k_{yp}$ lead to

$$E_p = E_n$$

$$\Rightarrow E_{cn} + E_{kn} = E_{vp} - E_{kp}$$

$$\frac{\hbar^2 k_{xn}^2}{2}\left(\frac{1}{m_{xn}} + \frac{1}{m_{xp}}\right) + \frac{\hbar^2 k_{yn}^2}{2}\left(\frac{1}{m_{yn}} + \frac{1}{m_{yp}}\right) + \frac{\hbar^2 k_{zn}^2}{2m_{zn}} = (E_{vp} - E_{cn}) - \frac{\hbar^2 k_{zp}^2}{2m_{zp}} \quad (S6)$$

Here we define $E_{max} = E_{vp} - E_{cn}$. Electrons emerging on the *p*-side must have a non-zero momentum in the *z* direction (not conserved due to the application of electric field), so $k_{zp}^2 \geq 0$, which will result in

$$\frac{\hbar^2 k_{xn}^2}{2}\left(\frac{1}{m_{xn}} + \frac{1}{m_{xp}}\right) + \frac{\hbar^2 k_{yn}^2}{2}\left(\frac{1}{m_{yn}} + \frac{1}{m_{yp}}\right) + \frac{\hbar^2 k_{zn}^2}{2m_{zn}} \leq (E_{vp} - E_{cn}) \quad (S7)$$

This condition sets the upper limit of the integration. For simplicity, we define $1/m_x^* = 1/m_{xn} + 1/m_{xp}$ and $1/m_y^* = 1/m_{yn} + 1/m_{yp}$. Therefore, the final expression of the tunneling current density is obtained as

$$J_t = \frac{4qg_sg_v\hbar}{(2\pi)^3 m_{zn}} \int_0^{\sqrt{\frac{2m_x^* E_{max}}{\hbar^2}}} dk_{xn} \int_0^{\sqrt{\frac{2m_y^*\left(E_{max} - \frac{\hbar^2 k_{xn}^2}{2m_x^*}\right)}{\hbar^2}}} dk_{yn} \int_0^{\sqrt{\frac{2m_{zn}\left(E_{max} - \frac{\hbar^2 k_{xn}^2}{2m_x^*} - \frac{\hbar^2 k_{yn}^2}{2m_y^*}\right)}{\hbar^2}}} dk_{zn} k_{zn}$$

$$(F_n - F_p) \; \exp\left(-\frac{4t\sqrt{2m_b^*}}{3q(V-V_i)\hbar}\left[(E_b - E_{cn} - E_{kn})^{\frac{3}{2}} - (E_b - E_{cn} - E_{kn} - q(V - V_i))^{\frac{3}{2}}\right]\right) \quad (S8)$$

The calculated tunneling current using this expression is shown in Figs. 2-4 of the main text by using four fitting parameters: $t$, $E_b$, $E_{fn} - E_{cn}$ and $E_{vp} - E_{fp}$. An exceptional agreement capturing both the reverse-bias and forward bias NDR characteristics is achieved, validating the model. The parameters used are reasonable, as discussed in the main text. The resultant curves exhibit NDR at forward bias voltages and the tunneling current rapidly increases at reverse bias, exhibiting a pronounced backward-diode behavior. Both these features are fingerprints of electron tunneling.

In additional to the tunneling component, the total current measured in at large forward bias voltages arise due to two components: the excess current, and the thermonic diffusion current.[2] The band-to-band tunnel current is dominant at low biases $V_{ds} < V_{valley}$, (here $V_{valley}$ is defined as the voltage corresponding to the minimum valley current) and the excess current becomes significant near $V_{valley}$. At higher positive biases thermoionic diffusion currents dominate. The excess current has observed in traditional Esaki diodes is attributed to carrier tunneling by way of energy states within the forbidden gap and scattering by photons, phonons or other electrons[2].

## 2) Additional electrical and TEM characterization of the reported device

The back gated I-V characteristics of the individual BP and SnSe$_2$ flakes (Fig. S1a) confirm that the BP flakes are all *p*-type and the SnSe$_2$ are all *n*-type as evidenced by the opposite directions of the gate modulation. Since both types of flakes are relatively thick and heavily doped, the current modulation is weak. Fig. S1b shows the stability of the measured $I_d$-$V_{ds}$ curves with NDR over a characterization period of months.

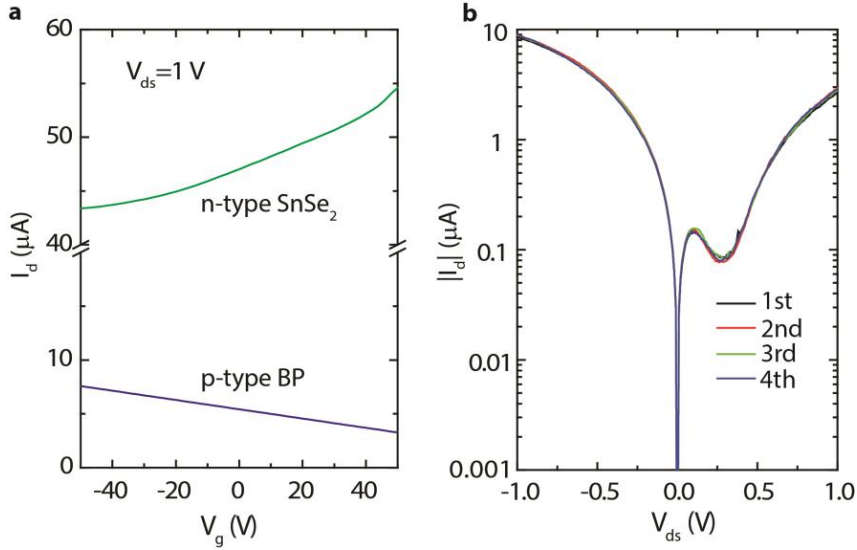

***Figure S1.*** *a, Back gated SnSe$_2$ and BP confirming the doping type of each flake material. b, Consecutive sweeps of the BP/SnSe$_2$ Esaki Diode showing stable $I_d$-$V_{ds}$ characteristics.*

Temperature dependent two-terminal $I_d$-$V_{ds}$ characteristics of individual BP and SnSe$_2$ flakes (not the Esaki diode) are shown in Fig. S2. The I-Vs of SnSe$_2$ remained linear at low temperatures. However, the I-Vs of BP became increasingly nonlinear with decreasing temperature. This indicates that the voltage dropped across the contacts and the access regions $I_d(T) * R_c(I_d, T)$ is responsible for the slight deviations leading to higher peak and valley voltages at lower temperatures in Fig 4a compared to the model in Fig 4b of the main text.

By comparing the modeled $I_d$-$V_{ds}$ curves with the experimental results and recognizing $V_{ds} = V_c(I_d, T) + V_j(I_d, T) = I_d(T) * R_c(I_d, T) + V_j(I_d, T)$, where $V_j(I_d, T)$ is the net voltage drop over the junction, we can deduce the external resistance $R_c(I_d, T)$. The results are shown in the inset of Fig. S2c. $R_c$ decreases with increasing temperature and bias voltages; and $R_c$ is in the range of ~M$\Omega$ at 80 K, which is very close to that of the total channel and contact resistance of BP extracted from Fig. S2b. This further confirms that the observed voltage shift in $I_d$-$V_{ds}$ at lower temperatures is a result of increased contact/access resistances.

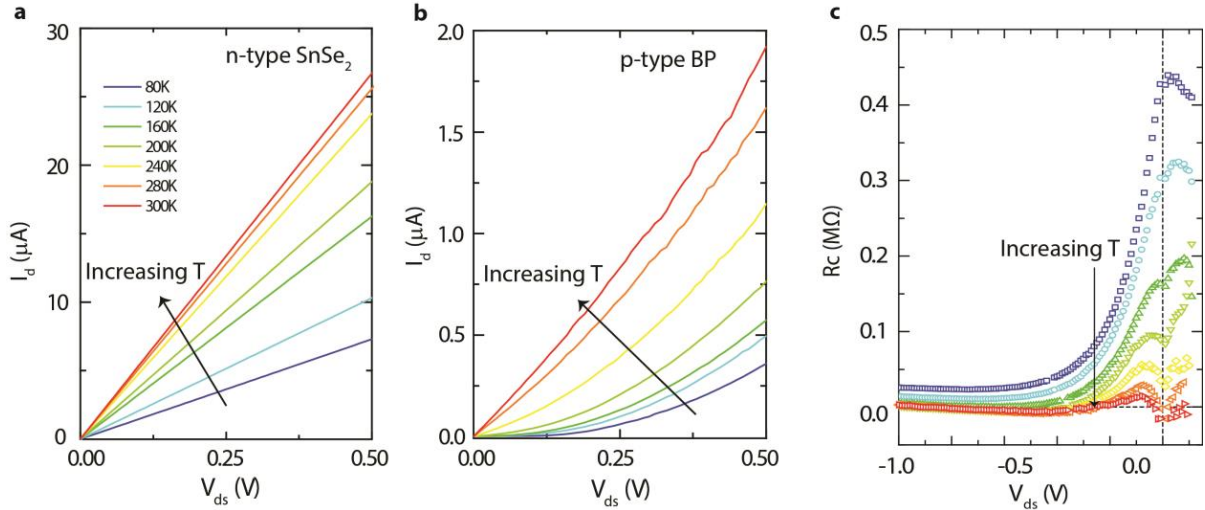

***Figure S2.*** ***a, b,*** *$I_d$-$V_{ds}$ characteristics of individual SnSe$_2$ and BP flakes over a temperature range of 80 K to 300 K.* ***c,*** *The extracted $R_c$ by comparing the theoretical and experimental results. $R_c$ includes both the metal/semiconductor contact resistances and BP/SnSe$_2$ access region resistances.*

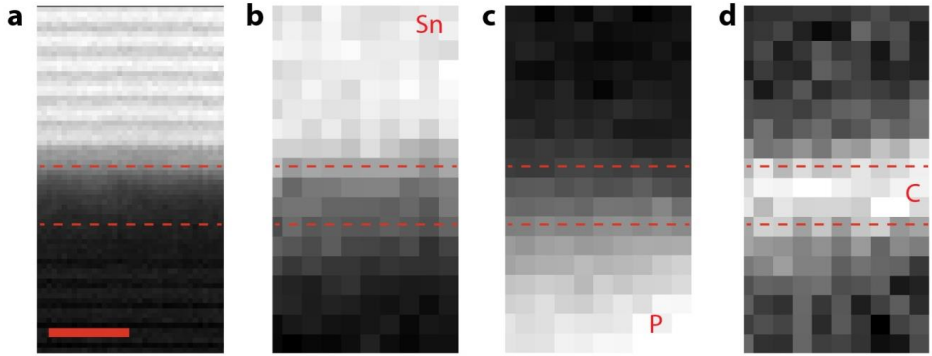

***Figure S3.*** ***a.*** *HAADF-STEM images of the interface between SnSe$_2$ and BP. An amorphous gap ranging from 1.2 nm to 2 nm was observed. The scale bar is 2 nm.* ***b-d*** *EELS composition maps showing the observed amorphous layer is composed of C, SnSe$_2$ and BP. BP layers have been reported to be difficult to image by TEM even for BP encapsulated between BN layers[3]; the crystal distortion was attributed to a few minutes of exposure to air between FIB and STEM imaging. In this work we did not observe significant electron beam damage at 100 kV during STEM imaging or in the EELS mapping.*

### 3) Reproducibility of vdW Esaki diodes with NDR

In Fig. S4 another device with stable NDR behavior is shown. The back-gated flake channel $I_d$-$V_{ds}$ confirm that $SnSe_2$ is n-type and BP is p-type (Fig. S4b), and at room temperature, decent ohmic contacts are obtained on both $SnSe_2$ and BP (Fig. S4c).

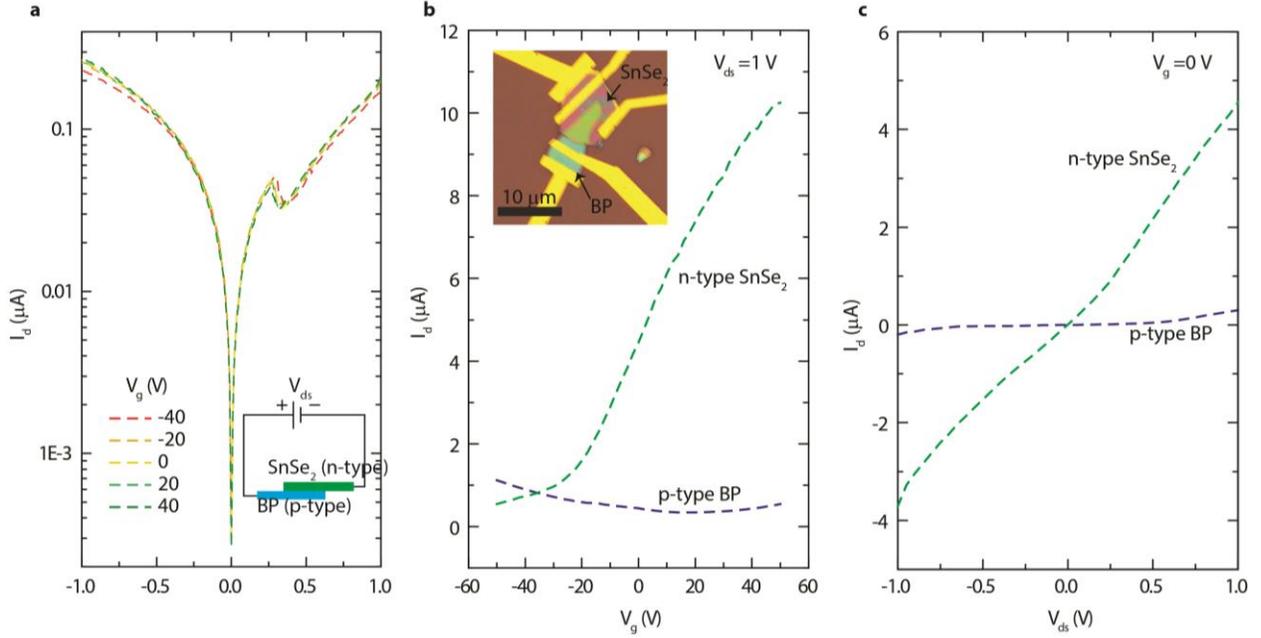

*Figure S4. Reproducibility of vdW BP/SnSe$_2$ Esaki diodes. **a,** $I_d$-$V_{ds}$ characteristics at several back gate biases. **b,** Drain current of individual flakes as a function of the back gate voltage. Inset shows the optical image of the measured device. **c.** Current-voltage characteristics of individual SnSe$_2$ and BP flakes showing the metal contacts are ohmic at RT.*

### 4) Observation of backward diode behavior without NDR

Most likely due to variations in the doping concentrations and interface quality of the BP/SnSe$_2$ vdW *p-n* diodes, some fabricated devices exhibit a backward diode behavior *without* NDR. In Fig. S5, we show one representative $I_d$-$V_{ds}$ characteristics of such a backward diode device without NDR. The backward diode behavior is already distinct from a normal *pn* junction: the reverse rectifying behavior with higher current at reserve bias than at forward bias is already a direct proof of electron tunneling. Whether NDR appears in a backward diode or not depends on the doping concentrations. At moderate doping densities, the Fermi levels in the n- and p-sides do not enter the bands, and therefore no NDR can be observed in the forward bias, as may be inferred from the energy band diagrams in Figs 2(e,f) of the main text. But these diodes still show backward-diode behavior, as Fig 2(d) proves. At higher degenerate doping densities the diodes show NDR under forward bias. Another possibility is that the NDR region is completely overwhelmed by the high excess current in these devices.

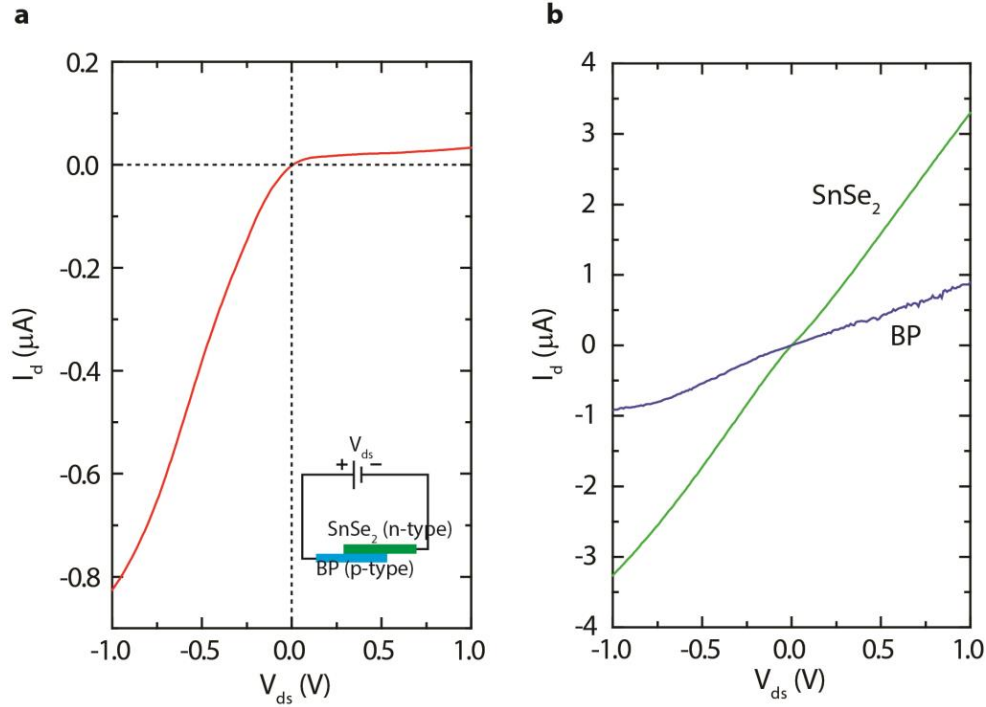

*Figure S5 A representative p-n BP/SnSe$_2$ backward diode. **a,*** $I_d$-$V_{ds}$ *characteristics over the heterojunction.* ***b,*** *Contacts on BP, SnSe$_2$ are ohmic.*

5) **Deduction of band alignment from photocurrent measurements**

From the photocurrent measurements shown in Fig. 5 in the main text, we can further confirm the broken-gap band alignment inferred from the observation of NDR. As reported in the literature[4,5] as well as in our own measurements shown in Fig. S1, exfoliated BP is p-doped and SnSe$_2$ is n-doped. If the Fermi level in BP is close to its valence band edge and that in SnSe$_2$ is close to its conduction band edge near the heterojunction, Figs. S6a-c illustrate the possible band diagrams assuming either broken-gap or staggered band alignment between BP and SnSe$_2$. Since the photocurrent measurements suggest an *accumulation* of carriers near the BP/SnSe$_2$ junction, a broken-gap alignment shown in Fig. S6a is thus confirmed. We exclude the situations shown in Figs. S6b-c because depletion of the charge carriers near the junction would lead to the measured I-V curves under illumination to the 4$^{th}$ quadrant. This is not observed in our experiment – we see the I-V to move to the 2$^{nd}$ quadrant.   It is also possible to have carrier accumulation for a staggered band alignment, as shown in Fig. S6d, only if the Fermi level in BP is close to the midgap or conduction band. But this contradicts the observed heavy p-type doping concentration in BP in our experiments.  Therefore, we conclude that BP and SnSe$_2$ form a broken gap band alignment.

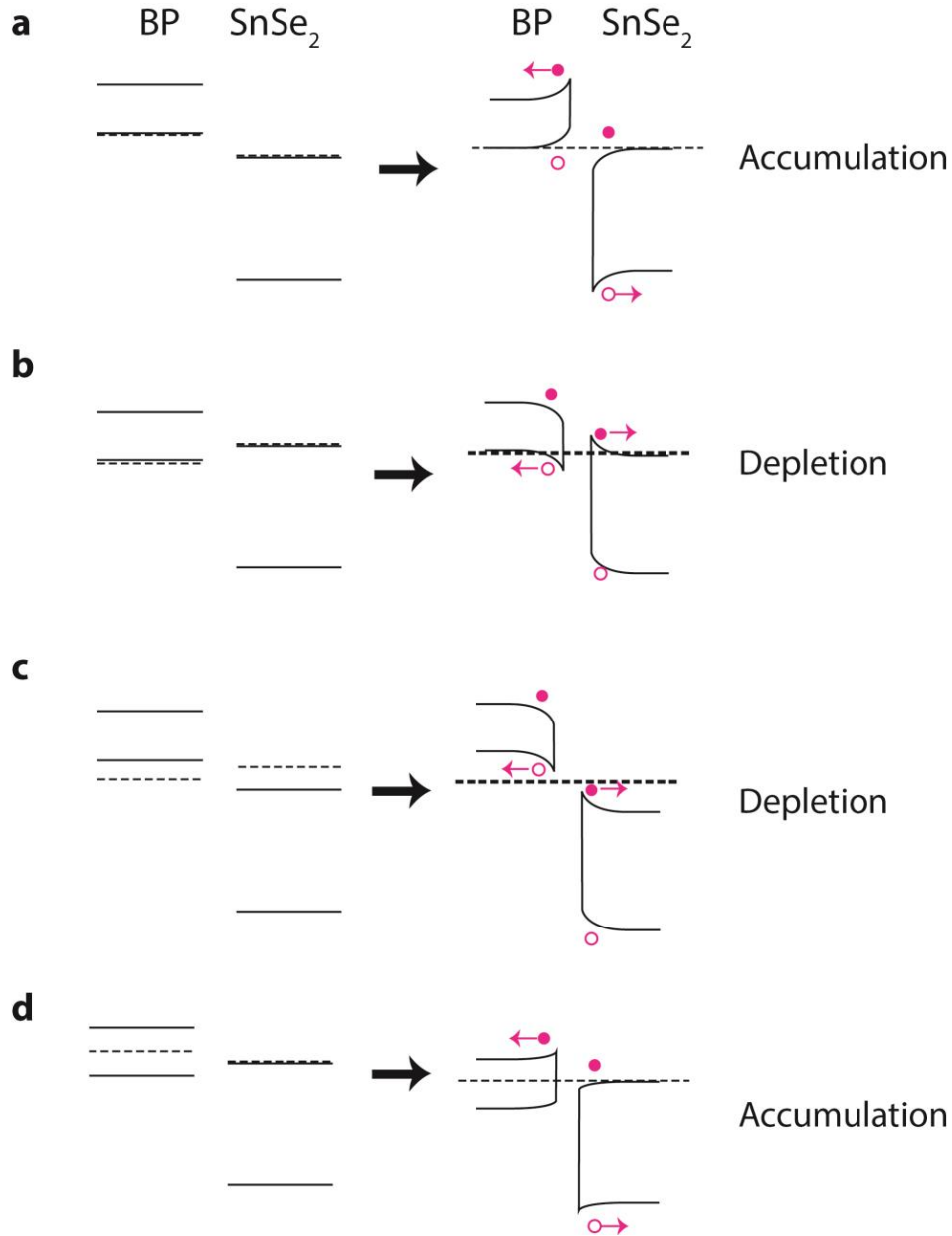

*Figure S6 Possible band alignments between BP and SnSe$_2$. **a, b** and **c,** Band alignments based on the fact that BP is heavily p-doped and SnSe$_2$ is heavily n-doped. Only the broken-gap band alignment in **a** supports carrier accumulation near the junction. **d,** Band alignment if the Fermi level in BP is*

*close to the mid-gap (lightly doped) or conduction band (n-type), which is not observed in our experiments.*